\documentclass[runningheads]{svmult}

\usepackage{makeidx}   % allows index generation
\usepackage{graphicx}  % standard LaTeX graphics tool
\usepackage{subeqnar}  % subnumbers individual equations
\usepackage{multicol}  % used for the two-column index
\usepackage{physprbb}  % modified textarea for proceedings,
\makeindex             % used for the subject index
\begin{document}

\title*{Stars and Fundamental 
Physics\footnote{Prepared for the Proceedings of the
ESO-CERN-ESA Symposium on Astronomy, Cosmology and Fundamental Physics 
(4--7 March 2002, Garching, Germany).}}

\toctitle{Stars and Fundamental Physics}
\titlerunning{Stars and Fundamental Physics}

\author{Georg G.~Raffelt}
\authorrunning{Georg G.~Raffelt}
\institute{Max-Planck-Institut f\"ur Physik 
(Werner-Heisenberg-Institut), F\"ohringer Ring 6,
80805 M\"unchen, Germany}

\maketitle

\begin{abstract}
Stars are powerful sources for weakly interacting particles that are
produced by nuclear or plasma processes in their hot interior.  These
fluxes can be used for direct measurements (e.g.\ solar or supernova
neutrinos) or the back-reaction on the star can be used to derive
limits on new particles. We discuss two examples of current interest,
the search for solar axions by the CAST experiment at CERN and
stellar-evolution limits on the size of putative large extra
dimensions.
\end{abstract}

%%%%%%%%%%%%%%%%%%%%%%%%%%%%%%%%%%%%%%%%%%%%%%%%%%%%%%%%%%%%%%%%%%%%%%%
%% Introduction %%%%%%%%%%%%%%%%%%%%%%%%%%%%%%%%%%%%%%%%%%%%%%%%%%%%%%%
%%%%%%%%%%%%%%%%%%%%%%%%%%%%%%%%%%%%%%%%%%%%%%%%%%%%%%%%%%%%%%%%%%%%%%%

\section{Introduction}

Astrophysics and cosmology provide a natural testing ground for
virtually any new idea in the area of elementary particle physics.
Usually one may first think of the early universe or perhaps
high-energy cosmic rays when searching for astrophysical arguments in
favor or against a new particle-physics model.  However, there are a
number of interesting cases where the low energies available in stars
are quite sufficient for rather useful and restrictive tests of
high-energy physics phenomena.

The basic idea is very simple. Stars are powerful sources for weakly
interacting particles such as neutrinos, gravitons, hypothetical
axions, and other new particles that can be produced by nuclear
reactions or by thermal processes in the hot stellar interior.  The
solar neutrino flux is now routinely measured with such precision that
compelling evidence for neutrino oscillations has accumulated. The
measured neutrino burst from supernova (SN) 1987A has been used to
derive many useful limits. Even when the particle flux can not be
measured directly, the absence of visible decay products, notably x-
or $\gamma$-rays, can provide important information. The properties of
stars themselves would change if they lost too much energy into a new
channel.  This ``energy-loss argument'' has been widely used to
constrain a long list of particles and particle properties. All of
this has been extensively reviewed~\cite{Raffelt1996,Raffelt:1999tx}
and is now widely appreciated among particle
physicists~\cite{Groom:in}.

Therefore, instead of reviewing once more the general ideas I will
rather focus on two topical examples of current interest that nicely
illustrate the overall methods. One is the search for solar axions by
the CAST experiment at CERN (Sec.~2). The other is the possibility
that space-time has large extra dimensions. This hypothesis predicts a
``tower'' of graviton modes that can be produced in stars, notably in
SN cores or neutron stars. The most restrictive limits on the size of
the extra dimensions arises from the astrophysical arguments presented
in Sec.~3.  A brief summary and outlook is given in Sec.~4.

%%%%%%%%%%%%%%%%%%%%%%%%%%%%%%%%%%%%%%%%%%%%%%%%%%%%%%%%%%%%%%%%%%%%%%%
%% Axion-Like Particles %%%%%%%%%%%%%%%%%%%%%%%%%%%%%%%%%%%%%%%%%%%%%%%
%%%%%%%%%%%%%%%%%%%%%%%%%%%%%%%%%%%%%%%%%%%%%%%%%%%%%%%%%%%%%%%%%%%%%%%

\section{Axion-Like Particles}

Axions are hypothetical particles that are predicted in the context of
a theoretical scheme to solve the CP problem of strong
interactions~\cite{Kim:ax,Cheng:1987gp}.  This is the problem that
quantum chromodynamics (QCD) ought to violate the CP symmetry in that
the neutron should have a large electric dipole moment, contrary to
experimental evidence. This observation can be explained by a new
symmetry, the Peccei-Quinn symmetry, that is spontaneously broken at
some large energy scale $f_a$, the Peccei-Quinn scale or axion decay
constant. Axions are the ``almost'' Nambu-Goldstone bosons of this new
symmetry and as such nearly massless.

Phenomenologically one should think of axions as the neutral pion's
little brother.  Model-dependent details aside, the axion's mass and
couplings are given by those of the $\pi^0$, scaled with $f_\pi/f_a$
where $f_\pi=93~{\rm MeV}$ is the pion decay constant.  The axion
decay constant $f_a$ is a free parameter and thus can be very large.
Therefore, axions can be very light and very weakly interacting even
though they are fundamentally a QCD phenomenon. 

There are other plausible solutions of the strong CP problem.
However, the Peccei-Quinn approach is particularly elegant and
predicts something new---in the guise of axions it provides a handle
for a possible experimental verification.  Moreover, axions can play
the role of the cosmic cold dark matter~\cite{Kolb:vq}.  Therefore,
two fundamental problems would be solved by the existence of one new
particle.

The experimental search for axions has focused on their predicted
interaction with the electromagnetic field that would be of the form
\begin{equation}
{\cal L}_{a\gamma\gamma}=
{\textstyle\frac{1}{4}} g_{a\gamma\gamma} 
F_{\mu\nu}\widetilde F^{\mu\nu}\,a=
-g_{a\gamma\gamma}{\bf E}\cdot{\bf B}\,a\,,
\end{equation}
where $F$ is the electromagnetic field-strength tensor, $\widetilde F$
its dual, and ${\bf E}$ and ${\bf B}$ the electric and magnetic
fields, respectively. The coupling strength is
\begin{equation}
g_{a\gamma\gamma}
=\frac{\alpha}{2\pi f_a}\,C_\gamma,
\qquad C_\gamma=\frac{E}{N}-1.92\pm0.08\,,
\end{equation}
where $E/N$ is the ratio of the electromagnetic and color anomalies, a
model-dependent ratio of small integers.  One popular case is the DFSZ
model where $E/N=8/3$, another the KSVS model where $E/N=0$, but there
are more general examples~\cite{Kim:1998va}.

\begin{figure}[b]
\begin{center}
\includegraphics[width=1.0\textwidth]{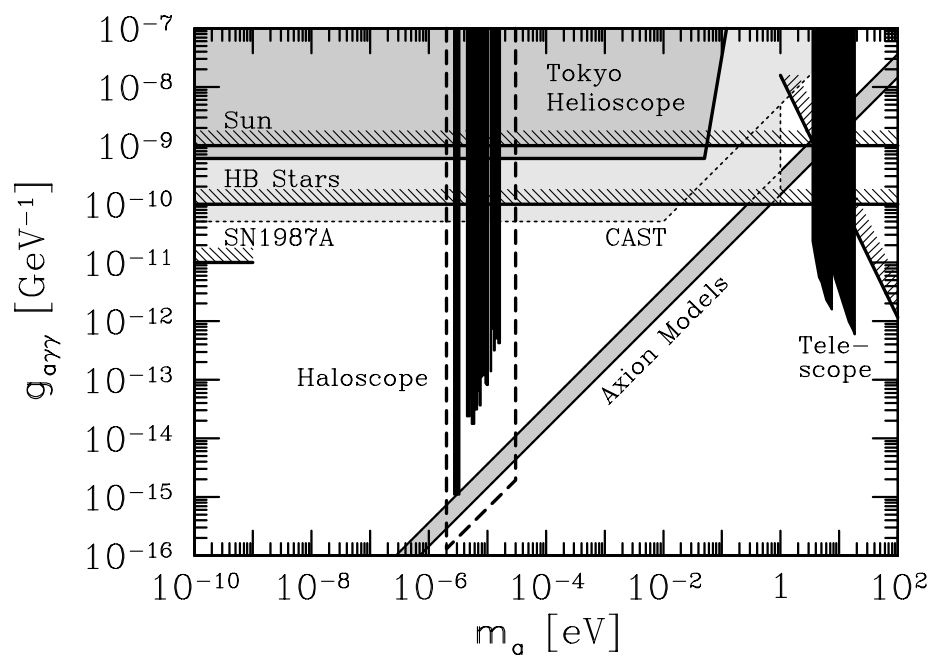}
\end{center}
\caption[]{Limits on the axion-photon coupling $g_{a\gamma\gamma}$ as
a function of axion mass $m_a$. The limits apply to any axion-like
particle except for the ``haloscope'' search which assumes that axions
are the galactic dark matter; the dotted region marks the projected
sensitivity range of the ongoing full-scale searches.  Limits for
higher masses than shown here are reviewed in
Ref.~\cite{Masso:1997ru}.  The light-grey region marks the foreseen
CAST sensitivity.}
\label{fig:axgam}
\end{figure}

Assuming $m_a=0.60~{\rm eV}\times 10^7~{\rm GeV}/f_a$ for the axion
mass, Fig.~\ref{fig:axgam} shows $g_{a\gamma\gamma}$ as a function of
$m_a$. The diagonal band marked ``Axion Models'' is somewhat
arbitrarily delimited by the DFSZ and KSVZ models.  The role of axions
or axion-like particles is frequently assessed in the full
two-dimensional $g_{a\gamma\gamma}$-$m_a$-space rather than the narrow
band defined by conventional axion models, although this band remains
the best-motivated location in this parameter space.

The electromagnetic interaction allows for the two-photon decay
$a\to\gamma\gamma$ with a rate $\Gamma_{\rm decay}=g_{a\gamma\gamma}^2
m_a^3/64\pi$.  This process is very slow if the axion mass is small
and the coupling strength is weak. Therefore, it is more promising to
consider the analogous process where one of the photons is virtual,
i.e.\ an external electric or magnetic field.  The
$a\leftrightarrow\gamma$ conversion in the presence of an external $E$
or $B$ field is known as the Primakoff process; it was first
considered for neutral pions half a century ago~\cite{Primakoff1951}.

If axions are the galactic dark matter, they can be detected in the
laboratory by the ``haloscope'' technique~\cite{Sikivie:ip}.  One
places a tunable high-Q microwave cavity in a strong magnetic field
and measures the power output. If the resonance frequency matches
$m_a$, the Primakoff-conversion can produce a measurable signal. Two
pilot experiments~\cite{Wuensch:1989sa,Hagmann:tj} and a first
full-scale search~\cite{Asztalos:2001jk,Asztalos:2001tf} exclude a
range of coupling strength shown in Fig.~\ref{fig:axgam} that is
marked ``Haloscope.''  The new generation of experiments in
Livermore~\cite{Asztalos:2001tf} and Kyoto~\cite{Yamamoto:2000si}
should cover the dashed area in Fig.~\ref{fig:axgam}, perhaps leading
to the discovery of axion dark matter.

In a different region of masses and couplings axions are detectable
with a related technique called the
``helioscope''~\cite{Sikivie:ip,vanBibber:1988ge}.  Thermal photons in
the solar interior convert to axions by the Primakoff process in the
microscopic electric fields of charged particles, producing a solar
axion flux which peaks at energies of a few keV.  If one views the Sun
through a long dipole magnet, the axions partially back-convert into
photons and become visible as x-rays at the far end of the magnet. A
dedicated search for this effect by the Tokyo Axion
Helioscope~\cite{Inoue:2002qy} excludes the dark-grey region in
Fig.~\ref{fig:axgam}.

\begin{figure}[b]
\begin{center}
\includegraphics[width=1.0\textwidth]{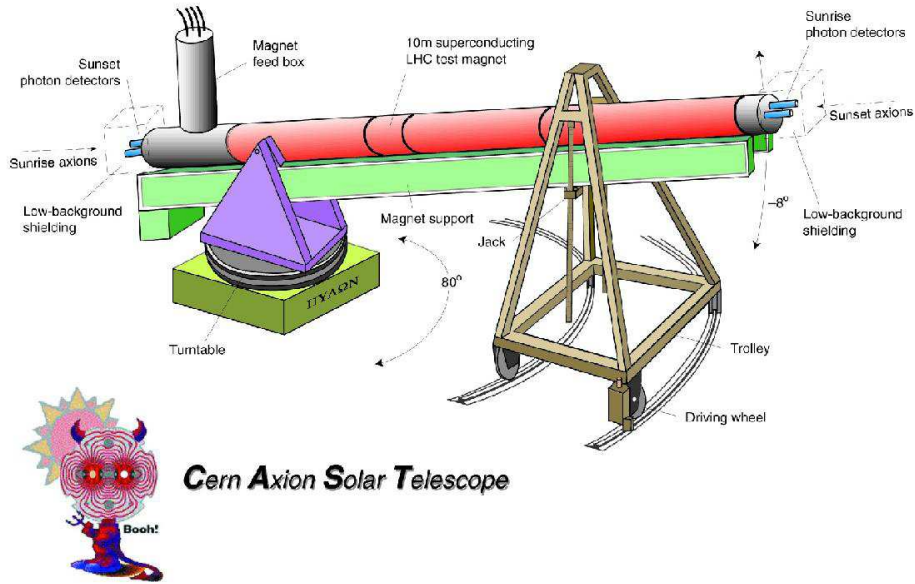}
\end{center}
\caption[]{Schematic view of the CAST experiment at CERN.}
\label{fig:cast}
\end{figure}

The conversion rate in the helioscope scales quadratically with the
length $L$ and field-strength $B$ of the conversion region. Therefore,
one can do much better in the new CAST project at CERN where a
de-commissioned LHC test magnet is used as a ``magnetic telescope'' to
search for solar axions~\cite{Zioutas:1998cc,CAST}. Mounted on a
movable platform (Fig.~\ref{fig:cast}) allowing $\pm40^\circ$
horizontal and $\pm5^\circ$ vertical tracking, this instrument can
achieve about 33 full days of alignment with the Sun per year.  If we
express the coupling strength as
$g_{a\gamma\gamma}=g_{10}\,10^{-10}~{\rm GeV}^{-1}$, the solar axion
flux at Earth is $g_{10}^2\,3.5\times10^{11}~{\rm cm^{-2}~s^{-1}}$.
The conversion probability in the magnet is
$g_{10}^2\,1.8\times10^{-17} (B/8.4\,{\rm T})^2(L/10\,{\rm m})^2$. For
the two magnet bores with a cross section of $2\times 14~{\rm cm}^2$
we thus expect an x-ray event rate of $15\,g_{10}^4$ per day of
exposure time.

In order to reach the sensitivity shown as a light-grey area in
Fig.~\ref{fig:axgam} one needs to make great efforts to suppress
background counts. One way is to focus the x-rays to a small detector
region. Specifically, an engineering model for the seven x-ray
telescopes of the Abrixas satellite has become available for this
purpose and has been tested to be in good working condition.  CAST
should be able to take first data shortly, i.e.\ in the summer or fall
of 2002.

Figure~\ref{fig:axgam} shows a loss of sensitivity for about
$m_a>10~{\rm meV}$. The axion-photon conversion should be pictured as
a phenomenon similar to neutrino oscillations~\cite{Raffelt:1987im}.
For larger $m_a$ the oscillation length becomes shorter than the
magnet and the effective mixing angle is suppressed. This ``momentum
mismatch'' between axions and photons can be overcome by giving the
photons a refractive mass by virtue of a low-$Z$ gas such as
helium. This approach was successfully employed in the Tokyo
Helioscope~\cite{Inoue:2002qy} and will be used in CAST as well.  We
may extend the sensitivity range to larger masses as shown in
Fig.~\ref{fig:axgam} and in the neighborhood of $m_a\sim 1~{\rm eV}$
actually bite into the parameter range of conventional axion models.

At somewhat larger masses axions are already ruled out by a telescope
search for spectral lines from $a\to\gamma\gamma$ decay in galaxy
clusters \cite{Bershady:1990sw}.  In the few-eV mass range axions
would have been in thermal equilibrium in the early universe and
contribute a small hot-dark matter component.

If we use the Sun as an axion source we must be sure that our
sensitivity range is not excluded by an excessive modification of
stellar properties by the axionic energy loss. An observable
modification of the solar p-mode frequencies excludes
$g_{a\gamma\gamma}$ values above the horizontal line in
Fig.~\ref{fig:axgam} marked ``Sun''~\cite{Schlattl:1998fz}.
Significantly smaller couplings are excluded because the energy-loss
of horizontal-branch (HB) stars would shorten their helium-burning
lifetime, reducing the relative number of HB stars observed in
globular clusters~\cite{Raffelt1996,Raffelt:1999tx}; see the
horizontal line in Fig.~\ref{fig:axgam} marked ``HB Stars.''  The CAST
experiment advances into uncharted territory.

For very small axion masses, however, the CAST sensitivity range is
already excluded by an argument involving SN~1987A. Axions would have
been produced in the hot SN core by the Primakoff effect, and then
would have back-converted into $\gamma$-rays in the galactic magnetic
field. The non-observation of a $\gamma$-ray burst in the SMM
instrument in coincidence with the observed SN~1987A neutrinos
excludes $g_{a\gamma\gamma}$ values above the line marked SN~1987A
\cite{Brockway:1996yr,Grifols:1996id}. This limit applies only for
about $m_a<10^{-9}~{\rm eV}$; for larger masses the conversion is
suppressed by the mass difference relative to photons.

The magnetically induced transition from photons to axion-like
particles in intergalactic space has been proposed as a mechanism that
would make distant photon sources look dimmer, with important
consequences for the interpretation of the SN Ia Hubble diagram
\cite{Csaki:2001yk,Csaki:2001jk,Erlich:2001iq,%
Deffayet:2001pc,Mortsell:2002dd}.  The relevant masses are very small,
again to avoid suppressing the transition by a large axion-photon mass
difference.  Therefore, the relevant $g_{a\gamma\gamma}$ range is
limited by the SN~1987A argument and thus falls outside the CAST
sensitivity range.

%%%%%%%%%%%%%%%%%%%%%%%%%%%%%%%%%%%%%%%%%%%%%%%%%%%%%%%%%%%%%%%%%%%%%%%
%% Large Extra Dimensions %%%%%%%%%%%%%%%%%%%%%%%%%%%%%%%%%%%%%%%%%%%%%
%%%%%%%%%%%%%%%%%%%%%%%%%%%%%%%%%%%%%%%%%%%%%%%%%%%%%%%%%%%%%%%%%%%%%%%

\section{Large Extra Dimensions}

The Planck scale of about $10^{19}$~GeV, relevant for gravitation, is
very much larger than the electroweak scale of about 1~TeV of the
particle-physics standard model.  A radical new approach to solving
this notorious hierarchy problem holds that there could be large extra
dimensions, the main idea being that the standard-model fields are
confined to a 3+1 dimensional brane embedded in a higher dimensional
bulk where only gravity is allowed to
propagate~\cite{add98,Antoniadis:1998ig,add99,Han:1999sg,grw99}.  This
concept immediately puts stringent constraints on the size of the
extra dimensions because Newton's law holds at any scale which has
thus far been observed, i.e.\ down to about 1~mm.  Extra dimensions
can only appear at a smaller scale.

Following common practice the new dimensions are taken to form an
$n$-torus of the same radius $R$ in each direction.  The Planck scale
of the full higher dimensional space, $M_{{\rm P},n+4}$, can be
related to the normal Planck scale,\footnote{Some authors define the
Planck mass as $M_{\rm P,4}=1.22\times10^{19}~{\rm GeV}/
(8\pi)^{1/2}=2.4\times10^{18}~{\rm GeV}$.  Limits on $M_{{\rm P},n+4}$
in this system of units have been reviewed in
Ref.~\cite{Uehara:2002yv}.}  $M_{{\rm P},4}=1.22\times10^{19}~{\rm
GeV}$, by Gauss' law~\cite{add98}
\begin{equation}
M_{{\rm P},4}^2 = R^n M_{{\rm P},n+4}^{n+2}.
\end{equation}
Therefore, if $R$ is large then $M_{{\rm P},n+4}$ can be much smaller
than $M_{{\rm P},4}$.  If this scenario is to solve the hierarchy
problem then $M_{{\rm P},n+4}$ must be close to the electroweak scale,
i.e.\ $M_{{\rm P},n+4} < 10$--100~TeV.  This requirement already
excludes $n=1$ because $M_{{\rm P},n+4} \simeq 100$~TeV corresponds to
$R \simeq 10^{8}$~cm.  However, $n \geq 2$ remains possible, and
particularly for $n=2$ there is the intriguing perspective that the
extra dimensions could be accessible to experiments probing gravity at
scales below 1~mm.

The most restrictive limits on $M\equiv M_{{\rm P},n+4}$ obtain from
supernovae and neutron stars.  The first example is the SN~1987A
energy-loss argument.  If large extra dimensions exist, the usual 4D
graviton is complemented by a tower of Kaluza-Klein (KK) states,
corresponding to new phase space in the bulk.  These KK gravitons
would be emitted from the SN core after collapse by nucleon
bremsstrahlung $N+N\to N+N+{\rm KK}$.  The KK gravitons interact with
the strength of ordinary gravitons and thus are not trapped in the SN
core.  However, this energy-loss channel can compete with neutrino
cooling because of the large multiplicity of KK modes and shorten the
observable
signal~\cite{Cullen:1999hc,SN1987A,Hanhart:2001er,Hanhart:2001fx}.
This argument has led to the tight bound $R < 0.66~\mu$m ($M >
31$~TeV) for $n=2$ and $R < 0.8$~nm ($M > 2.75$~TeV) for
$n=3$~\cite{Hanhart:2001fx}.

The KK gravitons emitted by all core-collapse SNe over the age of the
universe produce a cosmological background of these particles.  Later
they decay into all standard-model particles which are kinematically
allowed; for the relatively low-mass modes produced by a SN the only
channels are ${\rm KK} \to 2\gamma$, $e^+e^-$ and $\nu \bar\nu$.  The
relevant decay rates are $\tau_{2\gamma} = \frac{1}{2}\tau_{e^+e^-} =
\tau_{\nu \bar\nu} \simeq 6 \times 10^{9}~{\rm yr} \, (m/100~{\rm
MeV})^{-3}$ \cite{Han:1999sg}.  Therefore, over the age of the
universe a significant fraction of the produced KK modes has decayed
into photons, contributing to the diffuse cosmic $\gamma$-ray
background observed by EGRET.  This argument implies that if the
number of extra dimensions $n=2$ or 3, their radius $R$ must be about
a factor of 10 smaller than implied by the SN~1987A cooling limit,
i.e.\ for $n=2$ one finds $R < 0.9 \times 10^{-4}$~mm or $M \geq
84$~TeV.  For $n=3$ the new limit is $R < 0.19 \times 10^{-6}$~mm or
$M > 7$~TeV \cite{Hannestad:2001jv}.

This, however, is not the end of the story. We later realized that the
KK gravitons emitted by the SN core will stay gravitationally trapped
because most of them are produced near their kinematical threshold,
i.e.\ with barely relativistic velocities~\cite{Hannestad:2001xi}.
Therefore, every neutron star is surrounded by a halo of KK gravitons
which is dark except for the decays into $\simeq 100~{\rm MeV}$
neutrinos, $e^+e^-$ pairs and $\gamma$-rays.  In principle, this
radiation can be directly observed. Conversely, the non-observation
allows one to set stringent limits. In addition, the radiation
impinges on the neutron star, keeping it hot, above the observed
temperature in some cases such as the pulsar PSR J0953+0755. One
obtains the limit $M>1680~{\rm TeV}$ for $n=2$ and $M>60~{\rm TeV}$
for $n=3$.  In view of these limits one expects that if large extra
dimensions solve the hierarchy problem, their number $n$ should
probably exceed~4.

Similar arguments can be applied to other particles than gravitons
that may exist and may be able to propagate in the bulk of the
larger-dimensional space. The hypothetical majorons are one case in
point~\cite{Hannestad:2002ff}.

Of course, there are loop holes to such limits. The size of the extra
dimensions need not be equal, or there can be other than toroidal
compactifications. The KK gravitons may be able to decay fast into
invisible channels, and so forth. However, our main point is that
straightforward astrophysical arguments lead to non-trivial and
restrictive limits on the structure of this new theory.

%%%%%%%%%%%%%%%%%%%%%%%%%%%%%%%%%%%%%%%%%%%%%%%%%%%%%%%%%%%%%%%%%%%%%%%
%% Summary %%%%%%%%%%%%%%%%%%%%%%%%%%%%%%%%%%%%%%%%%%%%%%%%%%%%%%%%%%%%
%%%%%%%%%%%%%%%%%%%%%%%%%%%%%%%%%%%%%%%%%%%%%%%%%%%%%%%%%%%%%%%%%%%%%%%

\section{Summary and Outlook}

Stars continue to provide some of the most restrictive limits on new
particle-physics ideas. The much-discussed hypothesis that our
space-time has extra dimensions that are compactified on the
sub-millimeter scale is a recent case in point.  In addition to
deriving limits, there are opportunities for new discoveries. The CAST
experiment at CERN searching for solar axions will have a sensitivity
range that for the first time pushes beyond stellar-evolution limits
and thus has a realistic chance of actually finding axion-like
particles emitted by the Sun.

In future the observation of solar neutrinos will continue to provide
valuable information. The ongoing efforts in neutrino physics
virtually guarantee that large detectors will operate for many years
to come; even a megatonne detector may be built to search for proton
decay and to perform precision measurements at laboratory neutrino
beams. Therefore, chances are that one will measure the neutrino burst
from a galactic supernova, providing high-statistics information both
on the SN event and a host of information of particle physics
interest.

The recent excitement about the possible discovery of strange-matter
stars \cite{Seife2002}, even though not conclusive, illustrates that
compact stars offer one of the few opportunities to discover the true
ground state of nuclear matter.

Astroparticle physics is now an established research activity at the
interface between inner space and outer space.  The physics and
observationd of stellar objects continue to offer a number of
intriguing opportunities in this multi-faceted and interdisciplinary
field.

%%%%%%%%%%%%%%%%%%%%%%%%%%%%%%%%%%%%%%%%%%%%%%%%%%%%%%%%%%%%%%%%%%%%%%%
%% Acknowledgments %%%%%%%%%%%%%%%%%%%%%%%%%%%%%%%%%%%%%%%%%%%%%%%%%%%%
%%%%%%%%%%%%%%%%%%%%%%%%%%%%%%%%%%%%%%%%%%%%%%%%%%%%%%%%%%%%%%%%%%%%%%%

\section*{Acknowledgements}

This work was partly supported by the Deut\-sche
For\-schungs\-ge\-mein\-schaft under grant No.\ SFB 375 and the ESF
network Neutrino Astrophysics.

%%%%%%%%%%%%%%%%%%%%%%%%%%%%%%%%%%%%%%%%%%%%%%%%%%%%%%%%%%%%%%%%%%%%%%%
%% References %%%%%%%%%%%%%%%%%%%%%%%%%%%%%%%%%%%%%%%%%%%%%%%%%%%%%%%%%
%%%%%%%%%%%%%%%%%%%%%%%%%%%%%%%%%%%%%%%%%%%%%%%%%%%%%%%%%%%%%%%%%%%%%%%


\begin{thebibliography}{99}

\bibitem{Raffelt1996}
G.G.~Raffelt: {\em Stars as Laboratories for Fundamental Physics\/}
(Chicago University Press, Chicago, 1996)

\bibitem{Raffelt:1999tx}
G.G.~Raffelt:
``Particle physics from stars,''
Ann. Rev. Nucl. Part. Sci.  {\bf 49}, 163 (1999).
%[arXiv:hep-ph/9903472].
%%CITATION = HEP-PH 9903472;%%

\bibitem{Groom:in}
D.E.~Groom et al.  (Particle Data Group):
``Review of particle physics,''
Eur. Phys. J. C {\bf 15}, 1 (2000).
%%CITATION = EPHJA,C15,1;%%

\bibitem{Kim:ax}
J.E.~Kim:
``Light pseudoscalars, particle physics and cosmology,''
Phys. Rept. {\bf 150}, 1 (1987).
%%CITATION = PRPLC,150,1;%%

\bibitem{Cheng:1987gp}
H.Y.~Cheng:
``The strong CP problem revisited,''
Phys. Rept. {\bf 158}, 1 (1988).
%%CITATION = PRPLC,158,1;%%

\bibitem{Kolb:vq}
E.W.~Kolb and M.S.~Turner:
{\em The Early Universe}
(Addison-Wesley, Redwood City, 1990).

\bibitem{Kim:1998va}
J.E.~Kim:
``Constraints on very light axions from cavity experiments,''
Phys. Rev. D {\bf 58}, 055006 (1998).
%[arXiv:hep-ph/9802220].
%%CITATION = HEP-PH 9802220;%%

\bibitem{Primakoff1951}
H.~Primakoff:
``Photo-production of neutral mesons in nuclear electric fields 
and the mean life of the neutral meson,''
Phys. Rev. {\bf 81}, 899 (1951).

\bibitem{Sikivie:ip}
P.~Sikivie:
``Experimental tests of the `invisible' axion,''
Phys. Rev. Lett.  {\bf 51}, 1415 (1983),
Erratum ibid. {\bf 52}, 695 (1984).
%%CITATION = PRLTA,51,1415;%%

\bibitem{Wuensch:1989sa}
W.U.~Wuensch et al.:
``Results of a laboratory search for cosmic axions and other weakly
coupled light particles,''
Phys. Rev. D {\bf 40}, 3153 (1989).
%%CITATION = PHRVA,D40,3153;%%

\bibitem{Hagmann:tj}
C.~Hagmann, P.~Sikivie, N.S.~Sullivan and D.B.~Tanner:
``Results from a search for cosmic axions,''
Phys. Rev. D {\bf 42}, 1297 (1990).
%%CITATION = PHRVA,D42,1297;%%

\bibitem{Asztalos:2001jk}
S.J.~Asztalos et al.:
``Experimental constraints on the axion dark matter halo density,''
Astrophys. J.  {\bf 571}, L27 (2002).
%[arXiv:astro-ph/0104200].
%%CITATION = ASTRO-PH 0104200;%%

\bibitem{Asztalos:2001tf}
S.~Asztalos et al.:
``Large-scale microwave cavity search for dark-matter axions,''
Phys. Rev. D {\bf 64}, 092003 (2001).
%%CITATION = PHRVA,D64,092003;%%

\bibitem{Yamamoto:2000si}
K.~Yamamoto et al.:
``The Rydberg-atom-cavity axion search,''
hep-ph/0101200.
%%CITATION = HEP-PH 0101200;%%

\bibitem{Masso:1997ru}
E.~Masso and R.~Toldra:
``New constraints on a light spinless particle coupled to photons,''
Phys. Rev. D {\bf 55}, 7967 (1997).
%[arXiv:hep-ph/9702275].
%%CITATION = HEP-PH 9702275;%%

\bibitem{vanBibber:1988ge}
K.~van Bibber, P.M.~McIntyre, D.E.~Morris and G.G.~Raffelt:
``A practical laboratory detector for solar axions,''
Phys. Rev. D {\bf 39}, 2089 (1989).
%%CITATION = PHRVA,D39,2089;%%

\bibitem{Inoue:2002qy}
Y.~Inoue et al.:
``Search for sub-electronvolt solar axions using coherent conversion
  of axions into photons in magnetic field and gas helium,''
Phys. Lett. B {\bf 536}, 18 (2002).
%[arXiv:astro-ph/0204388].
%%CITATION = ASTRO-PH 0204388;%%

\bibitem{Zioutas:1998cc}
K.~Zioutas et al.:
``A decommissioned LHC model magnet as an axion telescope,''
Nucl. Instrum. Meth. A {\bf 425}, 482 (1999).
%[arXiv:astro-ph/9801176].
%%CITATION = ASTRO-PH 9801176;%%

\bibitem{CAST}
CERN Axion Solar Telescope homepage at
http://axnd02.cern.ch/CAST/

\bibitem{Raffelt:1987im}
G.~Raffelt and L.~Stodolsky:
``Mixing of the photon with low mass particles,''
Phys. Rev. D {\bf 37}, 1237 (1988).
%%CITATION = PHRVA,D37,1237;%%

\bibitem{Bershady:1990sw}
M.A.~Bershady, M.T.~Ressell and M.S.~Turner:
``Telescope search for multi-eV axions,''
Phys. Rev. Lett.  {\bf 66}, 1398 (1991).
%%CITATION = PRLTA,66,1398;%%

\bibitem{Schlattl:1998fz}
H.~Schlattl, A.~Weiss and G.~Raffelt:
``Helioseismological constraint on solar axion emission,''
Astropart. Phys.  {\bf 10}, 353 (1999).
%[arXiv:hep-ph/9807476].
%%CITATION = HEP-PH 9807476;%%

\bibitem{Brockway:1996yr}
J.W.~Brockway, E.D.~Carlson and G.G.~Raffelt:
``SN 1987A gamma-ray limits on the conversion of pseudoscalars,''
Phys. Lett. B {\bf 383}, 439 (1996).
%[arXiv:astro-ph/9605197].
%%CITATION = ASTRO-PH 9605197;%%

\bibitem{Grifols:1996id}
J.A.~Grifols, E.~Masso and R.~Toldra:
``Gamma rays from SN 1987A due to pseudoscalar conversion,''
Phys. Rev. Lett.  {\bf 77}, 2372 (1996).
%[arXiv:astro-ph/9606028].
%%CITATION = ASTRO-PH 9606028;%%

\bibitem{Csaki:2001yk}
C.~Csaki, N.~Kaloper and J.~Terning:
``Dimming supernovae without cosmic acceleration,''
Phys. Rev. Lett.  {\bf 88}, 161302 (2002).
%[arXiv:hep-ph/0111311].
%%CITATION = HEP-PH 0111311;%%

\bibitem{Csaki:2001jk}
C.~Csaki, N.~Kaloper and J.~Terning:
``Effects of the intergalactic plasma on supernova dimming via  
photon axion oscillations,''
Phys. Lett. B {\bf 535}, 33 (2002).
%[arXiv:hep-ph/0112212].
%%CITATION = HEP-PH 0112212;%%

\bibitem{Erlich:2001iq}
J.~Erlich and C.~Grojean,
``Supernovae as a probe of particle physics and cosmology,''
Phys.\ Rev.\ D {\bf 65} (2002) 123510.
%[arXiv:hep-ph/0111335].
%%CITATION = HEP-PH 0111335;%%

\bibitem{Deffayet:2001pc}
C.~Deffayet, D.~Harari, J.P.~Uzan and M.~Zaldarriaga:
``Dimming of supernovae by photon---pseudoscalar conversion and the
intergalactic plasma,''
hep-ph/0112118.
%%CITATION = HEP-PH 0112118;%%

\bibitem{Mortsell:2002dd}
E.~M\"ortsell, L.~Bergstr\"om and A.~Goobar:
``Photon axion oscillations and type Ia supernovae,''
astro-ph/0202153.
%%CITATION = ASTRO-PH 0202153;%%

%%%%%%%%%%%%%%%%%%%%%%%%%%%%%%%%%%%%%%%%%%%%%%%%%%%%%%%%%%%%%%%%%%%%%%%

\bibitem{add98}
N.~Arkani-Hamed, S.~Dimopoulos and G.~Dvali:
``The hierarchy problem and new dimensions at a millimeter,''
Phys. Lett. B {\bf 429}, 263 (1998).
%[hep-ph/9803315].
%%CITATION = HEP-PH 9803315;%%

\bibitem{Antoniadis:1998ig}
I.~Antoniadis, N.~Arkani-Hamed, S.~Dimopoulos and G. Dvali:
``New dimensions at a millimeter to a Fermi and superstrings 
at a TeV,''
Phys. Lett. B {\bf 436}, 257 (1998).
%[hep-ph/9804398].

\bibitem{add99}
N.~Arkani-Hamed, S.~Dimopoulos and G.~Dvali:
``Phenomenology, astrophysics and cosmology of theories with
sub-millimeter dimensions and TeV scale quantum gravity,''
Phys. Rev. D {\bf 59}, 086004 (1999).
%[hep-ph/9807344].
%%CITATION = HEP-PH 9807344;%%

\bibitem{Han:1999sg}
T.~Han, J.D.~Lykken and R.~Zhang:
Phys. Rev. D {\bf 59}, 105006 (1999).
%[hep-ph/9811350].
%%CITATION = HEP-PH 9811350;%%

\bibitem{grw99}
G.F.~Giudice, R.~Rattazzi and J.D.~Wells:
``Quantum gravity and extra dimensions at high-energy colliders,''
Nucl. Phys. B {\bf 544}, 3 (1999).
%[hep-ph/9811291].
%%CITATION = HEP-PH 9811291;%%

\bibitem{Uehara:2002yv}
Y.~Uehara:
``A mini-review of constraints on extra dimensions,''
hep-ph/0203244.
%%CITATION = HEP-PH 0203244;%%

\bibitem{Cullen:1999hc}
S.~Cullen and M.~Perelstein:
``SN~1987A constraints on large compact dimensions,''
Phys. Rev. Lett. {\bf 83}, 268 (1999).
%[hep-ph/9903422].
%%CITATION = HEP-PH 9903422;%%

\bibitem{SN1987A}
V.~Barger, T.~Han, C.~Kao and R.J.~Zhang:
``Astrophysical constraints on large extra dimensions,''
Phys. Lett. B {\bf 461}, 34 (1999).
%[hep-ph/9905474].
%%CITATION = HEP-PH 9905474;%%

\bibitem{Hanhart:2001er}
C.~Hanhart, D.R.~Phillips, S.~Reddy and M.J.~Savage:
``Extra dimensions, SN 1987A, and nucleon nucleon scattering data,''
Nucl. Phys. B {\bf 595}, 335 (2001).
%[nucl-th/0007016].
%%CITATION = NUCL-TH 0007016;%%

\bibitem{Hanhart:2001fx}
C.~Hanhart, J.A.~Pons, D.R.~Phillips and S.~Reddy:
``The likelihood of GODs' existence: Improving the SN 1987A 
  constraint on  the size of large compact dimensions,''
Phys. Lett. B {\bf 509}, 1 (2001).
%astro-ph/0102063.
%%CITATION = ASTRO-PH 0102063;%%

\bibitem{Hannestad:2001jv}
S.~Hannestad and G.G.~Raffelt:
``New supernova limit on large extra dimensions:
Bounds on Kaluza-Klein graviton production,''
Phys. Rev. Lett.  {\bf 87}, 051301 (2001).
%[arXiv:hep-ph/0103201].
%%CITATION = HEP-PH 0103201;%%

\bibitem{Hannestad:2001xi}
S.~Hannestad and G.G.~Raffelt:
``Stringent neutron-star limits on large extra dimensions,''
Phys. Rev. Lett.  {\bf 88}, 071301 (2002).
%[arXiv:hep-ph/0110067].
%%CITATION = HEP-PH 0110067;%%

\bibitem{Hannestad:2002ff}
S.~Hannestad, P.~Keranen and F.~Sannino:
``A supernova constraint on bulk majorons,''
hep-ph/0204231.
%%CITATION = HEP-PH 0204231;%%

%%%%%%%%%%%%%%%%%%%%%%%%%%%%%%%%%%%%%%%%%%%%%%%%%%%%%%%%%%%%%%%%%%%%%%%

\bibitem{Seife2002}
C.~Seife:
``If it quarks like a star, it must be \ldots\ strange?,''
Science {\bf 296}, 238 (2002).

\end{thebibliography}
\end{document}